# On combining features for single-channel robust speech recognition in reverberant environments

*José Novoa[1], Josué Fredes[1], Jorge Wuth[1], Fernando Huenupán[2], Richard M. Stern[3], Nestor Becerra Yoma[1]*

[1]Speech Process. and Transm. Lab., Elec. Eng. Dept., U.de Chile, Santiago, Chile.
[2]Dept. of Elec. Eng., Universidad de la Frontera, Temuco, Chile.
[3]ECE Dept. and Language Technologies Institute, CMU, Pittsburgh, PA 15213, USA.

`nbecerra@ing.uchile.cl`

**Abstract**

This paper addresses the combination of complementary parallel speech recognition systems to reduce the error rate of speech recognition systems operating in real highly-reverberant environments. First, the testing environment consists of recordings of speech in a calibrated real room with reverberation times from 0.47 to 1.77 seconds and speaker-to-microphone distances of 0.16 to 2.56 meters. We combined systems both at the level of the DNN outputs and at the level of the final ASR outputs. Second, recognition experiments with the reverb challenge are also reported. The results presented here show that the combination of features can lead to WER improvements between 7% and 18% with speech recorded in real reverberant environments. Also, the combination at DNN-output level is much more effective than at the system-output level. However, cascading both schemes can still lead to smaller reductions in WER.

## 1. Introduction

Many speech applications require that the user not be tethered to a close-talking microphone, such automatic meeting transcription, voice dialogue systems for devices in smart homes, and interaction with humanoid robots. In all of these cases the interactions are more intuitive, comfortable and effective if the user is able to interact with microphones on the device or at some third location, independent of the user. In many of these scenarios, the talker could be located several meters away from the microphone, and the received signal could be corrupted by interfering sounds, such as background noise and interfering speakers. In addition, speech in rooms is corrupted by the effects of reverberation caused by reflections of the speech from the surfaces of the room and the objects that are in it. The effects of reverberation are a major problem in distant-talking automatic speech recognition (ASR).

Reverberation (as well as background noise) decreases speech intelligibility and speech quality. This especially affects the performance of ASR systems, which are not as robust to reverberation as the human auditory system [1]. These performance degradations depend on the nature of the environment and make such systems less effective [2]; therefore, far-field speech recognition remains a challenge. One frequently-used measure of reverberation is the reverberation time (RT), which is defined as the time required for sound pressure level to decay by 60 dB. Offices and home environments typically have an RT from around 0.5 to 1.0 seconds, and longer RTs lead to greater reverberation distortion and greater degradation in ASR accuracy. Although RT is the most commonly-used general descriptor of reverberation, there are many other aspects of rooms that impact on speech intelligibility and ASR accuracy [3].

There are several reasons why reverberation is an especially difficult problem for automatic speech processing systems. First, the impulse response of typical rooms is much greater than the durations typically used for analysis frames in speech recognition or speaker identification systems, which means that the spectral colorations introduced cannot be ameliorated by cepstral mean normalization or other frame-based noise-removal algorithms and must be treated over a longer time scale (e.g. [4]). Many approaches based on adaptive noise cancellation that are quite successful in dealing with the effects of additive noise fail catastrophically in reverberant environments because the necessary assumption of statistical independence between the desired signal and the interfering source(s) is violated in reverberation, where the "distortion" consists of delayed and attenuated replications of the desired signal. Systems that are based on selective reconstruction based on time of arrival or relative intensity (*e.g.* [5]) fail because the reflections combine trigonometrically with the direct field and with each other to produce new amplitudes and phases that prevent the system from unambiguously inferring the direction of arrival of a sound source.

In recent years there has been increased interest in and attention to the development of systems that provide better robustness in reverberant environments, in part motivated by the REVERB challenge [6] and the IARPA ASpIRE Challenge [7], both of which included challenging data in a heterogeneous set of reverberant and noisy environments. Results of the challenges were reviewed in the REVERB Workshop held in conjunction with IEEE ICASSP 2014 and a special session in IEEE ASRU 2015, respectively. An excellent detailed analysis of the REVERB challenge may be found in [6]. It is noted in [6] that systems that were successful in the REVERB Challenge included a combination of enhancement at the waveform level and later, advanced acoustical modeling, typically using deep neural networks (DNNs), acoustic model adaptation, and occasionally system combination. Recent feature sets developed at least in part for reverberant environments include Damped Oscillator Coefficients (DOC) [8], Normalized Modulation Coefficients (NMC) [9], Modulation of Medium Duration Speech

Amplitudes (MMeDuSA) features [10], and Power Normalized Cepstral Coefficients (PNCC) [11]. Some of the compensation algorithms that have been developed to cope with reverberation include various missing-feature approaches (e.g. [12]), Suppression of Steady State and the Falling Edge (SSF) algorithm [13], nonnegative matrix factorization (e.g. [14]), and weighted prediction error (WPE) [15].

The major goals of this paper are to demonstrate that substantial improvements in recognition accuracy with speech recorded in real reverberant environments can be obtained by combining multiple ASR systems using complementary feature sets, trained using DNNs, and to analyze and compare combination techniques that lead to this efficiency. Finally, previous explorations of the use of feature combination in ASR have not directly addressed the problem of reverberant environments exhaustively.

## 2. Databases and procedures employed

In this section we describe the databases used and the details of our methods of experimentations, along with a small number of pilot results.

### 2.1. Databases for controlled experimentation using real reverberant speech

While the speech databases used for the REVERB and ASpIRE challenges provide a wealth of data with standardized evaluation results, we made use of our own publicly-available Highly-Reverberant Real Environments (HRRE) database for testing data because it incorporates calibrated and verified RTs and speaker-to-microphone distances. We believe that this controlled database is more suited for research purposes because it enables us to evaluate the impact of isolated subsets of the testing conditions. The training was accomplished using the augmented-data approach with environments that were similar to but not quite the same as the testing data. We describe these datasets below. It is worth highlighting that we also employed the Reverb challenge database to validate our findings.

#### 2.1.1. Testing with the HRRE database

We used the highly-reverberant real environments (HRRE) database developed by us for system testing. This database is described in detail in [16] and is publicly available for research purposes. The database consists of re-recorded utterances from the Aurora-4 clean evaluation set [17], which in turn were a subset of the evaluation-set utterances from the original Wall Street Journal (WSJ0) database [18]. The recording was performed in a real reverberant chamber, varying the speaker-to-microphone distance and the reverberation time. The speaker-to-microphone distances are 0.16, 0.32, 0.64, 1.28 and 2.56 meters. The room RTs are 0.47, 0.84, 1.27 and 1.77 seconds. Altogether, there are 20 combinations of speaker-to-microphone distances and RTs. Each combination consists of 330 testing utterances.

#### 2.1.2. Training for the HRRE database

We trained our systems using a multi-condition training database derived from the entire clean database of the WSJ0 SI-284 corpus, which consists of about 81 hours of clean speech. The training data were designed to be similar to but different from the test data. Each utterance of the clean database was convolved with three different simulated room impulse responses (RIRs) selected randomly from a list of 30,000 RIRs. The RIRs were simulated using the Room Impulse Response Generator [19]. The reverberation time (RT) values of the generated RIRs varied between 0.4 and 1.99 seconds. The dimensions of the virtual room used to generate each individual RIR were drawn from uniform distributions over the range of plus or minus 20 percent of the nominal values of 7.95 m length, 5.68 m width and 4.5 m height, which are the approximate dimensions of the non-rectangular room in which the test data were recorded. The speaker-to-microphone distance was drawn from a uniform distribution between 0.144 and 2.816 m. The speaker and microphone were placed in random locations at the room that were selected individually for a particular trial, with the constraints that both speaker and microphone were at least 1 m from any wall and between 1 m and 2 m from the floor. This randomization of the simulation parameters was implemented to reduce potential effects of artifacts caused by standing-wave phenomena in the rectangular shoebox-shaped room that RIR and other similar simulations based on the image method [20]. The resulting multi-condition database has a total duration of 325 hours, of which 25% are clean utterances, comparable to the Switchboard task [21].

### 2.2. Features used for ASR systems

Four feature sets were used in our experiments, which were selected to provide a degree of complementarity in representation. In each case the original signal was sampled at 16 kHz and windowed using 25-ms Hamming windows with 10 ms between frames. The feature sets were:

-Log Mel Filter Banks (MelFB): these features are perhaps the most widely used for DNN-HMM-based ASR systems. We used 40 Mel triangle weights as is common for 16-kHz speech, following [22].

-RASTA-Perceptual Linear Prediction (RPLP): these features are based on a more detailed auditory model than MelFB features, and the RASTA filter provides normalization along the time domain [23].

-Locally Normalized Filter Banks (LNFB): these features are also motivated by auditory perception and provide normalization with respect to frequency [24].

-Power-Normalized Cepstral Coefficients (PNCC): these physiologically-motivated features also contain components that provide robustness to reverberation and noise [11].

### 2.3. Description of the ASR system

We trained four DNN-HMM ASR systems in parallel using four different feature sets described above. WPE was applied to both training and testing. Each classifier was trained using the tri3a_dnn Kaldi Aurora-4 recipe in the Kaldi Speech Recognition Toolkit [25]. This recipe uses MFCC features and performs linear discriminant analysis (LDA) and maximum likelihood linear transformation (MLLT) to train a triphone system. As usual, a GMM-HMM recognizer was trained on clean data using the same recipe. This system is subsequently used to obtain clean forced alignments to the reverberant training data. The resultant alignments identify the senones used to train the DNNs [26]. The DNN architecture consists of seven hidden layers and 2048 units per layer. The input layer takes as input a context window of 11 frames, including the five frames before after the current frame. We used the Minimum Bayes Risk (MBR) decoding algorithm, with the standard 5K lexicon and trigram language model from the WSJ database.

## 3. Methods of system combination

In this section we describe the methods that we consider for system combination.

### 3.1. Background: complementarity of representations

The potential benefit that can be obtained from combining the computations of multiple ASR systems has been known for years, originally illuminated by the success of the NIST ROVER system [27] in combining the outputs of competing systems in DARPA speech recognition evaluations, almost invariantly producing results that were better than the best individual system. Indeed, system combination has proven to be highly effective in other domains besides speech processing as well (e.g. [28]). It was noted in an early prescient paper that system combination is most powerful when the systems to be combined are working from complementary "genetically different" information [29]. This is confirmed by multiple explorations of the topic in other domains (e.g. [30]).

### 3.2. Approaches to system combination

Most approaches to system combination fall into three broad classes. We can separate the speech recognition process into three principal stages: feature extraction, computation of phonetic probabilities by the acoustic models, and development of the recognition hypothesis using HMM decoding. In our ASR systems we use the four feature sets described in Section 2.2, DNNs to develop the probability of observing a given senone given the feature values, and the MBR algorithm for HMM decoding. System combination could take place at the level of the feature extraction, at the level of the DNN outputs, or at the level of the output hypotheses (e.g. [31]).

*Feature combination* is typically accomplished by concatenating the feature vectors of the systems to be combined and then applying techniques such as LDA, PCA, or the bottleneck DNN architecture to reduce the dimensionality of the resulting feature vector (e.g. [32]). *DNN-output combination* may be obtained by various operations on the outputs of the DNNs, which typically represent a posteriori probabilities or pseudo-log likelihoods (LLKs) of the senones given the data. Some of the methods used to combine probabilities or likelihoods over the years have included addition, weighted linear combination, multiplication, and computing the maximum of the scores (e.g. [31] [33]). *Output-level combination* is perhaps the most popular type of system combination, and describes different methods that combine parallel sources of information after the search procedure is completed. These methods include the well-known ROVER method [27], the hypothesis combination method [34], confusion network combination (CNC) [35], as well as various lattice-combination methods (e.g. [33]). For the experimental results described in this paper we will focus on DNN-output and output-level combinations.

### 3.3. DNN-output combination

Given $R$ ASR systems, each with its own ensemble of DNNs, the linear combination of scores is defined as:

$$\hat{m}(s,n) = \sum_{r=1}^{R} \omega_{r,s,n} \cdot m_r(s,n) \qquad (1)$$

where $\hat{m}$ is the combined score (log-likelihoods in this paper) representing the $s^{th}$ output (where s indicates the senone or state) for the $n^{th}$ input frame. Similarly, $m_r(s,n)$ is the score provided by the DNN representing the $r^{th}$ system, and $\omega_{r,s,n}$ is the corresponding weight. These weights could be dependent on or independent of state or frame under consideration. In this paper:

$$\omega_{r,s,n} = 1/R \qquad (2)$$

### 3.4. Output combination of multiple ASR systems

As noted above, output combination methods are those in which ASR systems are combined based on the scores complete hypotheses, which typically include the words that are hypothesized, their likelihood scores, and possibly their putative begin and end times. The seminal NIST ROVER method [27] combined scores based on "voting" across word identities. Subsequent improvements included the hypothesis combination approach [34] which, in addition, made use of word scores and begin/end times, and the lattice combination method [33], which exploited a richer description of the hypotheses. The Confusion Network (CNC) method [35] used dynamic programming to identify all paths in the "confusion network" that are relevant to a particular hypothesis. In our experiments we make use of the lattice-combination (LC) approach that is augmented by minimum Bayes risk (MBR) decoding [36] which provides slightly better performance than the similar CNC method without the MBR postprocessing.

## 4. Discussion

### 4.1. Results with the HRRE database

We begin our discussion of the experimental results by considering the results of the individual ASR systems for the HRRE test database. Word error rates (WERs) are shown as a function of RT averaged across speaker-microphone distances in Table 1. We observe that WER increases (unsurprisingly) as RT and speaker-to-microphone distance increase. We also note that the best-performing individual ASR system used MelFB features as input, i.e. 5.99%.

Table 2 shows the WERs achieved by combining MelFB with LNFB, PNCC and RPLP. We evaluated *DNN-output combination* with uniform weight as in (2) and *Output-level* combination with LC/MBR. When compared to MelFB, *DNN-output combination* and *Output-level combination* provided reduction in WER equal to 11.6% and 5.0%. This result strongly suggests that: different features can provide complementary information; and, to combine the DNN outputs can be a very effective scheme to combine features.

Table 1: *WER(%) obtained using MelFB, LNFB, PNCC and RPLP features with the HRRE database. WER is averaged across the speaker-to- microphone distances for each RT.*

| | RT [s] | | | | |
|---|---|---|---|---|---|
| **Feature** | **0.47** | **0.84** | **1.27** | **1.77** | **Avg.** |
| **MelFB** | 3.35 | 4.78 | 6.61 | 9.22 | 5.99 |
| **LNFB** | 3.68 | 5.15 | 6.70 | 9.64 | 6.29 |
| **PNCC** | 3.40 | 5.04 | 6.64 | 9.44 | 6.13 |
| **RPLP** | 4.33 | 6.78 | 8.62 | 11.92 | 7.91 |

Table 2: *Average WER(%) across all 20 HRRE testing data subsets with MelFB combined with the each one of the other feature methods.*

| Combination | MelFB-LNFB | MelFB-PNCC | MelFB-RPLP | Avg. |
|---|---|---|---|---|
| **DNN-output** | 5.24 | 5.31 | 5.33 | 5.29 |
| **LC/MBR** | 5.57 | 5.60 | 5.90 | 5.69 |

According to Table 3, the combination of all the features considered here (i.e. MelFB, LNFB, PNCC and RPLP) with *DNN-output combination* and *Output-level combination* provided reductions in WER as high as 17.5% and 11.9% when compared with MelFB, respectively. This result corroborates that *DNN-output combination* is a more suitable framework for feature combination than the *Output-level* one. When both combination schemes were employed (Fig. 1), the reduction in WER was 18%. This result suggests that the *DNN-output* method incorporates the combination at output level. For validation purposes, we trained four systems with MelFB using different random initializations and combine them with *DNN-output combination*, *Output-level combination* and both schemes in cascade (Fig. 1). Observe that the reductions in WER were hardly 5% when compared to MelFB.

### 4.2. Results with the Reverb challenge

To validate our results with the HRRE database, we ran experiments with the Reverb challenge. According to Table 4,

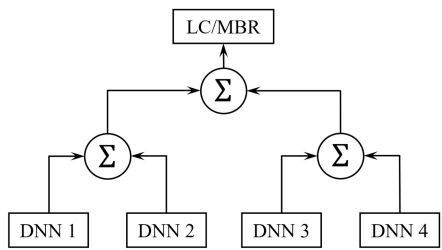

Figure 1: *DNN-output and System output combinations in cascade.*

Table 3: *Average WER(%) across all 20 HRRE testing data subsets obtained by combining all four feature methods. For comparison reasons, four systems with MelFB were also trained with different initializations.*

|  | MelFB –LNFB –PNCC –RPLP | 4 Init. MelFB |
|---|---|---|
| **DNN-output** | 4.94 | 5.69 |
| **LC/MBR** | 5.28 | 5.88 |
| **DNN output+ LC/MBR** | 4.91 | 5.67 |

Table 4: *WER(%) obtained with the Reverb Challenge data with systems trained by using MelFB, LNFB, PNCC, and RPLP. Also, MelFB was combined with LNFB, PNCC, and RPLP.*

|  | Sim data Avg. | Real data Avg. |
|---|---|---|
| **MelFB** | 7.2 | 22.5 |
| **LNFB** | 7.8 | 23.9 |
| **PNCC** | 8.5 | 23.8 |
| **RPLP** | 10.4 | 27.7 |
| **MelFB – LNFB** | 6.9 | 21.2 |
| **MelFB – PNCC** | 7.2 | 20.8 |
| **MelFB – RPLP** | 7.4 | 20.9 |

our base line system with MelFB provides competitive WERs when compared to [37], that won the 2014 Reverb challenge. The combination of features with at the DNN-output level could lead to a reduction of WER equal to 7% when compared to MelFB.

## 5. Conclusion

This paper addresses the combination of complementary parallel speech recognition systems in order to reduce the error rate of speech recognition systems operating in real highly-reverberant environments. We obtained substantial improvements in recognition accuracy, even though the original systems were well trained using state-of-the art deep learning approaches. The experiments in the present paper describe the recognition of speech that has been recorded in a controlled highly-reverberant real room and with the Reverb challenge database. The results presented here suggest that the combination of features can lead to WER improvements between 7% and 18% with speech recorded in real reverberant environments. Also, the combination at DNN-output level is much more effective than at the system-output level, although cascading both schemes can still lead to smaller reductions in WER. Evaluations with other deep learning architectures is proposed as future research.

## 6. Acknowledgements

The research reported here was funded by grants Conicyt-Fondecyt 1151306 and ONRG N°62909-17-1-2002.